\documentclass[twocolumn,twocolappendix]{aastex631}
\usepackage{amsmath}
\usepackage{comment}
\usepackage{multirow}

\begin{document}

\title{A TeV-based Determination of the Local Extragalactic Background Light and its Consistency with Galaxy Counts and Direct Measurements}

\author[0009-0004-9545-794X]{J. Baxter}
\affiliation{Institute for Cosmic Ray Research (ICRR), The University of Tokyo, Kashiwa, 277-8582 Chiba, Japan}
\affiliation{Visiting Researcher at IPARCOS and Department of EMFTEL, Universidad Complutense de Madrid, E-28040 Madrid, Spain}
\affiliation{Department of Physics and Astronomy, Clemson University, Clemson, SC, 29631, USA}
\correspondingauthor{J. Baxter}
\email{jrbaxte@clemson.edu}

\author[0000-0002-3433-4610]{A. Dom\'{i}nguez}
\affiliation{IPARCOS and Department of EMFTEL, Universidad Complutense de Madrid, E-28040 Madrid, Spain}

\author[0000-0001-5941-7933]{J.~D. Finke}
\affiliation{Space Science Division, Naval Research Laboratory, Washington, DC 20375-5352, USA}

\author[0000-0001-7405-9994]{A. Desai}
\affiliation{Dept. of Physics and Wisconsin IceCube Particle Astrophysics Center, University of Wisconsin-Madison, Madison, WI 53706, USA}

\author[0000-0002-6584-1703]{M. Ajello}
\affiliation{Department of Physics and Astronomy, Clemson University, Clemson, SC, 29631, USA}

\author[0000-0001-7796-8907]{A. Banerjee}
\affiliation{Department of Physics and Astronomy, Clemson University, Clemson, SC, 29631, USA}

\author[0000-0002-8028-0991]{Dieter Hartmann}
\affiliation{Department of Physics and Astronomy, Clemson University, Clemson, SC, 29631}

\author[0000-0001-7774-5308]{Vaidehi  S. Paliya}
\affiliation{Inter-University Centre for Astronomy and Astrophysics (IUCAA), SPPU Campus, Pune, 411007, Maharashtra, India}



\begin{abstract}

The extragalactic background light (EBL), the cumulative radiation from all extragalactic sources, traces galaxy formation and cosmic evolution. High-energy $\gamma$ rays attenuated via pair production with EBL photons are a powerful probe of the EBL. In this work, we use very-high-energy (VHE; $E_\gamma > 100\,\mathrm{GeV}$) $\gamma$ rays to measure the local EBL intensity and test its consistency with galaxy counts and direct measurements. Our analysis employs a sample of 268 spectra from 45 sources observed with Imaging Atmospheric Cherenkov telescopes. A model-dependent study shows seven EBL templates require only $\le 10\%$ rescaling to fit the observed $\gamma$-ray attenuation. The galaxy-count-anchored model gives the closest match. We then derive template-marginalized TeV optical depths from a representative model subset. We combine them with \textit{Fermi}-LAT GeV measurements to reconstruct the EBL at $z = 0$ using empirical and physically motivated models. The two reconstructions agree and follow the integrated galaxy light to within $2$--$3\,\mathrm{nW\,m^{-2}\,sr^{-1}}$ (typically $<25\%$) over $0.5$--$30\,\mu$m. Both are consistent with low-zodiacal-light observations, including outer solar system and dark cloud measurements. In contrast, the near-IR excess reported by IRTS and CIBER exceeds our reconstructed intensity by $3$--$5\sigma$, implying an additional $\gtrsim 5$--$10\,\mathrm{nW\,m^{-2}\,sr^{-1}}$ incompatible with the $\gamma$-ray optical depths. Combined with GeV constraints on EBL evolution to $z \simeq 4$, these TeV optical depths provide a VHE-anchored determination of the local EBL intensity. The agreement with galaxy counts and deep-space measurements indicates that known galaxy populations account for most of the optical and near-IR background, leaving limited room for an additional diffuse component.
\end{abstract}

\keywords{Gamma-ray astronomy (628) --- Observational cosmology (1146) --- Cosmic background radiation (317) --- Active galactic nuclei (16) --- Diffuse radiation (383)}


\section{Introduction} \label{sec:intro}
The extragalactic background light (EBL) represents the cumulative radiation produced by all stars and galaxies outside the Milky Way, spanning wavelengths from the ultraviolet (UV) to the far-infrared \citep[IR, e.g.,][]{1993ppc..book.....P, 1998ApJ...508...25H, 2013APh....43..112D, cooray2016, Hill_2018, Mattila_2019}. Potential diffuse contributions from non-stellar sources, such as accreting black holes, may also be present. Its spectral energy distribution and redshift evolution trace the history of star formation, galaxy assembly, Population~III contributions \citep[e.g.,][]{Matsumoto_2005, Inoue_2013}, the cosmic star formation history \citep[CSFH, e.g.,][]{2014ARA&A..52..415M}, and dust reprocessing. A precise characterization of the EBL is also essential for modeling extragalactic $\gamma$-ray propagation \citep[e.g.,][]{1994ApJ...423L...5A, biteau2022gamma, abdalla24}, interpreting blazar spectra \citep[e.g.,][]{dominguez2015spectral, paliya20}, and constraining cosmological parameters \citep{Dom_nguez_2013, Biteau_2015, dominguez19H0, Greaux_2024,2024MNRAS.527.4632D}.

The two most established observational approaches are direct photometric measurements and the integrated galaxy light (IGL). Direct measurements are limited by dominant foregrounds such as Zodiacal Light \citep{cooray2016}, though significant progress has come from New Horizons observations in the outer Solar System \citep{lauer2021new, lauer2022, symons2023, postman2024new} and the HST SKYSURF reanalysis \citep{windhorst22}. The IGL method has also improved substantially: \citet{Koushan_2021}, followed by more recent updates by \citet{2025arXiv250703412T}, revised the \citet{driver2016measurements} estimates, finding intensities higher by 5--15\% and reducing uncertainties below 10\% in most bands through deeper imaging and improved source separation. As a result, IGL estimates now better match recent direct measurements, helping resolve earlier discrepancies between the two approaches \citep[see, e.g.,][for discussions on this historical tension]{2013APh....43..112D, cooray2016, Mattila_2019}.

Gamma-ray attenuation offers an independent and increasingly precise route to constraining the EBL. High- and very-high-energy (VHE; $E_\gamma > 100\,\mathrm{GeV}$) $\gamma$ rays are attenuated via pair production with EBL photons \citep[e.g.,][]{gould1966opacity,1967PhRv..155.1408G}, imprinting an energy- and redshift-dependent signature on blazar spectra. Over the past decade, MAGIC, H.E.S.S., VERITAS, and the \textit{Fermi}-LAT have used GeV and TeV blazars to constrain the EBL, for example using over 700 sources in \citet{2018_science_fermi_ebl}. Most of these studies relied on earlier EBL models, which introduces systematic uncertainties.

\citet{2026Banerjee..EBL} complement our work by deriving updated GeV optical depths from 15~yr of \textit{Fermi}–LAT observations of 1576 blazars, reconstructing the EBL and its evolution to $z \simeq 4.3$. That study provides the most precise GeV–based EBL measurement to date and focuses on redshift evolution and its relation to the CSFH. Here we focus on TeV optical depths measured with ground-based Cherenkov telescopes. TeV $\gamma$ rays are attenuated through pair production with EBL photons with energy $\epsilon$, with the interaction cross-section peaking when $E_\gamma \, \epsilon \sim (m_e c^2)^2$, where $E_\gamma$ is the observed energy of the $\gamma$-ray photon. Specifically, $\gamma$ rays in the TeV range interact most efficiently with optical to near-infrared background light. This makes VHE observations highly sensitive to the EBL around and beyond its near-infrared peak, providing a crucial complement to GeV measurements that primarily probe the ultraviolet and optical regimes. Our goal is therefore to obtain an updated determination of the local EBL intensity and to place constraints on its low-redshift evolution.

In this work we pursue three goals, building primarily on \citet{2019ApJ...874L...7D}. First, using the STeVECat catalog \citep{greaux2024stevecat}, we test a broad set of recent EBL models in a model–dependent analysis and quantify their allowed rescaling. Second, we derive template-marginalized TeV optical depths $\tau(E_\gamma, z)$ by combining several EBL templates. Third, we combine these TeV optical depths with GeV $\tau(E_\gamma, z)$ measurements from \citet{2018_science_fermi_ebl} to reconstruct the EBL intensity and its low-redshift evolution. This is done using two different approaches, one motivated by a physical framework , and another empirically determined. We then compare the resulting $z = 0$ spectrum with IGL and direct measurements to assess the completeness of resolved galaxy populations.

The paper is organized as follows. Section \ref{sec:sample} describes the sample selection and data reduction. Section \ref{sec:method} details the methodology, including the gamma-ray attenuation framework, systematic uncertainties, and the EBL reconstruction models. Section \ref{sec:results} presents our results on the EBL density constraints, optical depths, and the reconstructed EBL spectrum. Finally, Section \ref{sec:discussion} discusses the consistency of our results with galaxy counts and direct measurements and summarizes our conclusions.

We adopt the concordance cosmology model of \( H_0 = 70 \, \mathrm{km \, s^{-1} \, Mpc^{-1}} \), \( \Omega_m = 0.3 \), and \( \Omega_\Lambda = 0.7 \), applying these parameters consistently throughout the analysis.

\section{Sample and data selection} \label{sec:sample}
Our analysis is based on the STeVECat catalog, a publicly available database that compiles 403 VHE $\gamma$-ray spectra from 73 extragalactic sources \citep{greaux2024stevecat}. These data were collected from observations by ground-based Cherenkov telescopes, including H.E.S.S., MAGIC, and VERITAS. It compiles archival spectra with at least two flux points from 173 journal publications, formatted to align with standards used in Gamma-Cat\footnote{An open data collection and source catalog for $\gamma$-ray astronomy, see \url{https://github.com/gammapy/gamma-cat}.} and VTSCat \citep{Acharyya_2023}. Metadata such as observation periods, livetime, source coordinates, and redshifts are also included, enabling systematic studies of the sources in the catalog.

Following the approach outlined by \citet{Greaux_2024}, we applied a redshift cut of 
$z>0.01$ and required a minimum of four flux points per spectrum to ensure robust measurements of EBL-induced absorption. Updating the approach of \citet{2019ApJ...874L...7D}, who reconstructed the EBL from 106 spectra compiled by \citet{Biteau_2015}, the present work leverages the larger STeVECat dataset together with more recent EBL models. After imposing $z>0.01$, requiring a minimum of four flux points per spectrum, and removing averaged spectra that duplicate flux-state-resolved observations of the same source, our final sample contains 268 spectra from 45 sources extending to $z=0.939$ (PKS 1441+25), considerably beyond the redshift range of \citet{2019ApJ...874L...7D}, which reached $z \approx 0.6$. The TeV optical depth measurements derived here constitute the VHE data set that complements the GeV optical depths used by \citet{2026Banerjee..EBL},  providing an independent, ground-based anchor for the EBL intensity at wavelengths near and beyond the near-IR peak ($\sim 1$--$10\,\mu\mathrm{m}$).

\section{Methodology}\label{sec:method}
In this section we describe the procedure used to derive the $\gamma$-ray optical depths and reconstruct the EBL intensity. We first outline the attenuation framework used to model the interaction between $\gamma$ rays and the EBL, then detail the strategy used to obtain template-marginalized optical depths. Finally, we present the reconstruction approaches adopted to infer the EBL spectrum and its evolution, and discuss the treatment of systematic uncertainties.

\subsection{Gamma-ray attenuation framework}\label{sec3.1}
Let the intrinsic energy spectrum of an extragalactic $\gamma$-ray source at redshift \( z \), defined as the spectrum that would be observed in the absence of EBL attenuation, be represented by \( \Phi_{\mathrm{int}}(E_{\gamma}) \). The observed spectrum, \( \Phi_{\mathrm{obs}}(E_{\gamma}) \), is then given by:

\begin{equation}
\label{abs_eq}
\Phi_{\mathrm{obs}}(E_{\gamma}) = \Phi_{\mathrm{int}}(E_{\gamma}) \exp\left(-\tau(E_{\gamma}, z)\right)
\end{equation}
where \( \tau(E_{\gamma}, z) \) denotes the expected optical depth with regard to the interaction between a $\gamma$-ray photon and an EBL photon, as predicted by a given EBL model.

Following the methodology first introduced by \citet{2012Sci...338.1190A}, and later employed by \citet{2018_science_fermi_ebl} and \citet{2019MNRAS.486.4233A}, we introduce an energy-independent scale factor, \( \alpha \), in the exponent of Eq.~(\ref{abs_eq}), modifying the absorption factor to the form \( \exp(-\alpha \tau) \). We considered four standard intrinsic spectral functions commonly used in the literature \citep[e.g.,][]{2012Sci...338.1190A, 2018_science_fermi_ebl}: power law, log-parabola, power law with an exponential cutoff, and log-parabola with an exponential cutoff. Following the approach of \citet{2019ApJ...874L...7D}, for each source and each EBL model, we first selected the preferred intrinsic spectral template based on the highest $\chi^2$ probability at the fiducial EBL normalization ($\alpha = 1$). 

Using this chosen intrinsic template, we performed a likelihood maximization fit for \( \alpha \) values ranging from 0 to 2.5, with a step size of 0.05. By multiplying the individual likelihoods across all sources within a given bin, we constructed a global profile likelihood for \( \alpha \). The global profile likelihood \( L(\alpha) \) yields \( \alpha_{\text{best}} \), where \( L \) is maximized, representing the best-fit EBL density relative to the model. According to Wilks' theorem \citep{Wilks:1938dza}, under the null hypothesis and asymptotic conditions, the test statistic \( TS = -2 \log \Lambda \), where \( \Lambda \) is the likelihood ratio \( L(\alpha) / L(\alpha_{\text{best}}) \), follows a \( \chi^2 \) distribution with one degree of freedom. This allows the determination of the 1$\sigma$ confidence interval, corresponding to the values of \( \alpha \) for which \( \Delta(-2 \log \Lambda) = 1 \).

The above procedure was performed for seven EBL models in total \citep[][]{Kneiske_2010, 2011MNRAS.410.2556D, Gilmore_2012, Inoue_2013, Franceschini_2017, 2021MNRAS.507.5144S, Finke_2022}. These were selected to encompass a diverse range of fundamental modeling methodologies: forward evolution via semi-analytical galaxy formation \citep{Gilmore_2012,Inoue_2013}, backward evolution based on low-$z$ galaxy counts \citep{Franceschini_2017}, cosmic star formation history-based modeling converting star formation rates to emission \citep{Finke_2022}, and empirical reconstructions from galaxy luminosity densities and spectral energy distributions \citep{2021MNRAS.507.5144S}. Additional models \citep{Kneiske_2010, 2011MNRAS.410.2556D} were included to provide historical baselines and bounding constraints.

\subsection{Optical depth measurement strategy}\label{sec3.2}

The measurement of the optical depth in this study follows the methodology employed in previous works \citep[][]{2018_science_fermi_ebl,2019ApJ...874L...7D}. Specifically, for each energy and redshift bin, we derived a stacked TS profile as a function of $\alpha$. While all seven EBL models listed were used in the model-dependent analysis of Section~\ref{sec3.1}, the optical depth measurement employs a subset of four. The four EBL models selected for this subset \citep{Inoue_2013, Franceschini_2017, 2021MNRAS.507.5144S, Finke_2022} were chosen a priori, specifically to span the distinct modern methodologies detailed in Section~\ref{sec3.1}. This diverse set helps reduce systematic biases tied to any single evolutionary prescription and provides a robust, template-marginalized estimate of the optical depth.

Within a given bin, the optical depth was determined as the average of the individual measurements obtained using these four models. This approach effectively marginalizes over the specific shape assumptions of any single template. While not strictly ``model-independent'' in the sense of a free-form fit, it reduces the dependence on any single evolutionary prescription by averaging over a diverse set of template shapes. To ensure a comprehensive treatment of uncertainties, the uncertainty in each bin was defined to encompass the full range of $\tau$ values across the four models. These uncertainties therefore represent a conservative systematic envelope combining statistical errors from the likelihood fit with the template-driven spread, rather than a strictly frequentist confidence interval. Internal checks using $\tau(E_\gamma, z)$ from individual templates, e.g.\ \citet{2021MNRAS.507.5144S}, yield consistent EBL reconstructions, confirming that the result is robust to the model selection.

\subsection{EBL Reconstruction Models}
To interpret the optical depth measurements derived in Section \ref{sec3.2}, we employ two reconstruction techniques: a physically motivated framework and an empirical reconstruction. The difference between the models is how the luminosity density $j(\lambda, z)$ is calculated. For both frameworks, we constrained the full set of free parameters using the ensemble MCMC sampler \texttt{emcee} \citep{emcee13}, assuming uniform (flat) priors for all parameters within their physically allowed ranges, and comparing the model $\tau_{\gamma\gamma}$ described below with the observational values discussed in Section \ref{sec3.2}. The EBL specific intensity at $z=0$ was then obtained by integrating the luminosity density over redshift:

\begin{equation}
\lambda I_{\lambda} = \frac{1}{4\pi} \int \mathrm{d}z \frac{\lambda j(\lambda/(1+z), z)}{(1+z)^2} \frac{\mathrm{d}\ell}{\mathrm{d}z}
\end{equation}
where $\lambda/(1+z)$ denotes the rest-frame wavelength corresponding to the observed wavelength, $\lambda$. To connect the reconstructed EBL photon density back to the observable $\gamma$-ray attenuation, we calculate the optical depth $\tau_{\gamma\gamma}(E_{\gamma}, z)$ assuming a flat $\Lambda$CDM cosmology as:

\begin{multline}
    \tau_{\gamma\gamma}(E_{\gamma}, z) = \int_0^z \mathrm{d}z' \frac{\mathrm{d}\ell}{\mathrm{d}z'} (z')
    \int_{-1}^{1} \mathrm{d}\mu \; \frac{1-\mu}{2} \\
    \int_{0}^{\infty} \mathrm{d}\epsilon \; n_{\mathrm{EBL}}(\epsilon, z')(1+z')^3 
    \sigma_{\gamma\gamma}(\beta', z') \Theta(\epsilon - \epsilon_{\text{th}}')
\end{multline}
where $\mu = 1 - \cos{\theta}$, with $\theta$ being the angle between the directions of the $\gamma$-ray photon and an EBL photon, $\frac{\mathrm{d}\ell}{\mathrm{d}z'} = \frac{c}{H_0 (1+z') \sqrt{\Omega_m(1+z')^3+\Omega_{\Lambda}}}$ is the line element as a function of redshift, $n_{\mathrm{EBL}}(\epsilon, z')$ is the number density of EBL photons with comoving energy $\epsilon$ at redshift $z'$, $\sigma_{\gamma\gamma}(\beta', z')$ is the Breit–Wheeler cross-section \citep{Breit_1934} of the interaction, and $\Theta(x)$ is the Heaviside step function that ensures the integration only includes values of $\epsilon$ above the threshold energy $\epsilon_{\text{th}}' = 2(m_e c^2)^2 / E_{\gamma}(1 - \mu)(1 + z') $, with $m_e$ being the mass of the electron\footnote{For the fundamental equations in this context, one can refer to \citet{cooray2016}.}.

\subsubsection{Physically Motivated Framework}
The physically motivated model builds upon the PEGASE \citep[Projet d'Etude des GAlaxies par Synthese Evolutive;][]{1997A&A...326..950F} spectral libraries, which compute synthetic spectra of evolving stellar populations. These are then convolved with a redshift-dependent star formation history while incorporating dust attenuation and the evolving metallicity of the universe. The star formation rate density is parameterized using the four-parameter function introduced by \citet{2014ARA&A..52..415M}, with free parameters $a_{s}$, $b_{s}$, $c_{s}$, and $d_{s}$. In addition, two parameters, $f_{1}$ and $f_{2}$, describe the fractions of absorbed starlight that are re-emitted in the infrared by two of the three dust components. We assume the initial mass function of \citet{baldry2003imf}. Based on this framework, the comoving luminosity density is computed, from which the EBL specific intensity and the $\gamma$-ray optical depth are derived \citep[see][]{Finke_2022}.

\subsubsection{Empirical Reconstruction}
In this reconstruction, the luminosity density, $j(\lambda, z)$, is modeled as a sum of log-normal distributions, each anchored at a fixed peak wavelength. This methodology, originally developed by \citet{2018_science_fermi_ebl} and further extended by \citet{2019ApJ...874L...7D}, provides a flexible and semi-empirical representation of the EBL spectrum. While previous studies have explored varying the number and position of these Gaussian components, we found that fixed logarithmic spacing provides sufficient flexibility to model the EBL without overfitting the current VHE dataset. Specifically, the comoving emissivity at wavelength $\lambda$ is given by:

\begin{equation}
j(\lambda) = \sum_{i} a_{i} \exp \left[ -\frac{(\log_{10} \lambda - \log_{10} \lambda_i)^2}{2\sigma^2} \right]
\end{equation}
where $a_{i}$ denotes the normalization of each component, $\lambda_{i}$ are fixed pivot wavelengths spaced logarithmically at $[0.16, 0.50, 1.6, 5.0, 16, 50, 160]\,\mu\mathrm{m}$, and $\sigma$ is the width of the distributions, fixed to 0.2. Each log-normal component is allowed to evolve independently with redshift according to:

\begin{equation}
j(\lambda_i, z) = j_0(\lambda_i) 
\begin{cases} 
\frac{(1+z)^{b_i}}{1+[(1+z)/c_i]^{d_i}} & \text{for } i \le 3 \\
(1+z)^{b_i} & \text{for } i > 3 
\end{cases}
\end{equation}
where $b_{i}$, $c_{i}$ and $d_{i}$ are free parameters describing the evolution of UV-to-optical components, while infrared components $(i > 3)$ are modeled with simple power-law evolution.

\subsection{Systematic uncertainties}
Two primary sources of systematic uncertainties were examined to evaluate their influence on the derived EBL scale factor. To simulate a maximally pessimistic scenario, a uniform $\pm15$ \% variation was applied to the energy scale across all samples, enabling an assessment of the EBL scale factor's sensitivity to deviations in energy calibration. In this context, we adopt this $\pm15$ \% energy-scale variation as a deliberately conservative choice. Typical energy-scale uncertainties reported by H.E.S.S., MAGIC, and VERITAS lie in the 10 to 15 \% range \citep{Aharonian2006_HESSperf, Park2015_VERITASperf, Aleksic2016_MAGICperf, Abe2023_magic_lst_joint}, and applying a coherent shift to all instruments maximizes the impact on the inferred EBL scaling. A full treatment of IRF and background systematics is not feasible with catalog-level flux points. Published instrument performance studies indicate that residual flux-level systematics at the relevant energies are typically at the $\sim 10$--$20$\% level and could mimic a coherent shift in the EBL normalization.

As a second source of systematic uncertainty, we excluded the power-law model from the intrinsic spectral model selection. This ensures that the results are not overly dependent on a specific choice of spectral shape, since the intrinsic spectral curvature is partially degenerate with EBL absorption. Specifically, 137 spectra initially preferred the power-law model when fitting at the fiducial normalization, justifying its exclusion as a robust test to capture model-driven systematic uncertainties. Our use of the remaining intrinsic spectral forms is intended to bracket this degeneracy.

Quantitatively, the energy-scale perturbation shifts the best-fit $\alpha$ by approximately $6\%$, while excluding the power-law model shifts it by about $4\%$. Combining these in quadrature yields a total systematic uncertainty of $\sigma_{\mathrm{sys}} \approx 0.07$, or $\sim7\%$ of the EBL scale factor. Since $\tau$ scales linearly with $\alpha$, this systematic propagates directly into the optical depth measurements.

\begin{figure*}[ht!]
\plotone{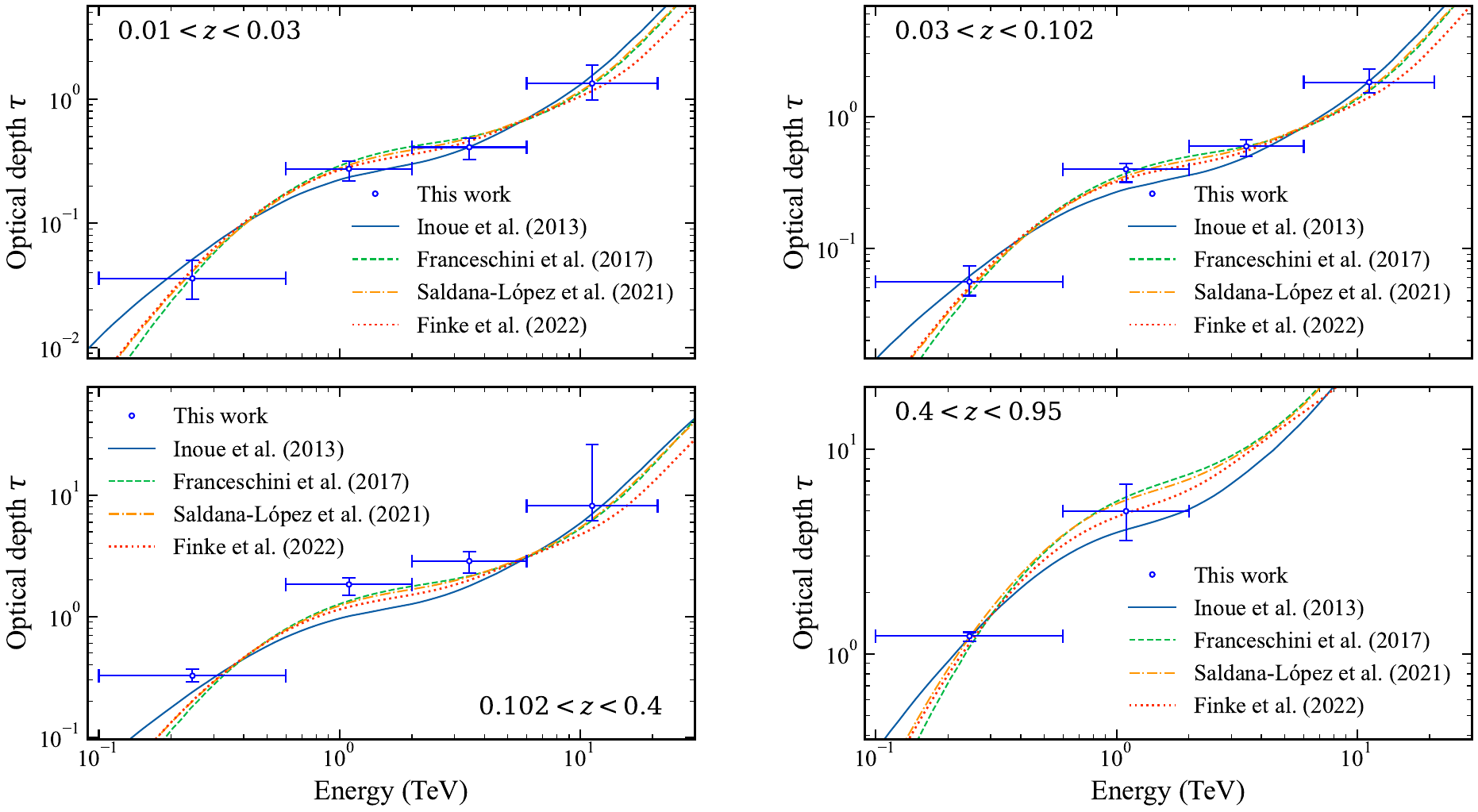}
\caption{Reconstructed optical depth $\tau$ as a function of $\gamma$-ray energy in four redshift bins: \textit{top left} (0.01 $< z <$ 0.03), \textit{top right} (0.03 $< z <$ 0.1), \textit{bottom left} (0.1 $< z <$ 0.4), and \textit{bottom right} (0.4 $< z <$ 0.95). Blue data points show the results of this work, with vertical error bars indicating $1\sigma$ statistical uncertainties and horizontal bars representing the energy binning. Predictions from EBL models are overplotted for comparison: \citet{Inoue_2013,Franceschini_2017,2021MNRAS.507.5144S, Finke_2022}.}
\label{fig:optical_depth}
\end{figure*}

\section{Results}\label{sec:results}

\subsection{TeV Optical Depths}
The results of our optical-depth measurements are presented in Fig.~\ref{fig:optical_depth} and tabulated in Table~\ref{tab:optical_depth}. A key outcome of this study is the ability to double the number of redshift bins, from 2 to 4, and extend their range from $z=0.6$ to $z\sim 1$, compared to \citet{2019ApJ...874L...7D}, made possible by the expanded dataset. Our TeV optical depth measurements are derived from a much larger spectral sample than \citet{2019ApJ...874L...7D}. In our EBL reconstruction studies, they are combined with the GeV optical depths from \citet{2018_science_fermi_ebl}, providing the VHE complement to the GeV $\tau(E_\gamma, z)$ used in \citet{2026Banerjee..EBL}.

\begin{table*}
\centering
\caption{Measured optical depths $\tau$ in different redshift and energy intervals.}
\label{tab:optical_depth}
\begin{tabular}{ccccc}
\hline\hline
Redshift & 0.10--0.60 & 0.60--2.00 & 2.00--6.00 & 6.00--21.00 \\
$(z)$ & (TeV) & (TeV) & (TeV) & (TeV) \\
\hline
$0.01$--$0.03$ & $0.04^{+0.01}_{-0.01}$ & $0.28^{+0.04}_{-0.06}$ & $0.41^{+0.08}_{-0.08}$ & $1.34^{+0.54}_{-0.36}$ \\
$0.03$--$0.102$ & $0.06^{+0.02}_{-0.01}$ & $0.40^{+0.04}_{-0.08}$ & $0.59^{+0.08}_{-0.09}$ & $1.82^{+0.48}_{-0.32}$ \\
$0.102$--$0.4$ & $0.32^{+0.04}_{-0.03}$ & $1.84^{+0.23}_{-0.34}$ & $2.86^{+0.56}_{-0.60}$ & $8.19^{+18.04}_{-2.00}$ \\
$0.4$--$0.95$ & $1.22^{+0.06}_{-0.07}$ & $4.98^{+1.74}_{-1.40}$ & -- & -- \\
\hline\hline
\end{tabular}
\end{table*}

\subsection{Evaluation of Standard EBL Models}
Table~\ref{tab:ebl_density} summarizes the best-fit EBL scale factors for all seven models tested. Among the EBL models considered, the \citet{Inoue_2013} model yields the lowest rejection significance for $\alpha=0$ ($15.4\sigma$, compared with $18$--$19\sigma$ for the other models). Two factors may contribute to this behavior. First, the predicted optical depth is systematically smaller in the energy and redshift range of our sample, which reduces the spectral impact of scaling the attenuation and broadens the likelihood profile. Second, the attenuation curve has a slightly smoother energy dependence around 1 to 3 TeV, which may increase the degeneracy with the intrinsic spectral curvature. The strongest rejection significance for the fiducial normalization $\alpha=1$ was found for the model by \citet{Kneiske_2010}, at the level of $2.4\sigma$. However, we should note that this model provides a strict lower limit of the EBL flux based on a forward evolution framework. As such, obtaining $\alpha>1$ is not only expected but also serves as a validation of the robustness of our analysis. 

\begin{table*}[ht]
\centering
\caption{Results of EBL density constraints (best-fit EBL scale factor $\alpha$)\label{tab:ebl_density}}
\begin{tabular}{lcccc}
\hline\hline
EBL Model & $\alpha_\text{best}$ & $R^{\sigma}(\alpha = 1)$ & $R^{\sigma}(\alpha = 0)$ & Test Statistic \\
\hline
\citet{Kneiske_2010}        & $1.12^{+0.06\;+0.07}_{-0.05\;-0.07}$      & $2.4\sigma$  & $18.4\sigma$ & 340 \\
\citet{2011MNRAS.410.2556D} & $0.91 \pm 0.05 \pm 0.05$                   & $1.8\sigma$  & $18.6\sigma$ & 346 \\
\citet{Gilmore_2012}        & $1.03^{+0.06\;+0.06}_{-0.08\;-0.05}$      & $0.8\sigma$  & $19.0\sigma$ & 369 \\
\citet{Inoue_2013}          & $1.05^{+0.12\;+0.04}_{-0.04\;-0.07}$      & $1.2\sigma$  & $15.4\sigma$ & 324 \\
\citet{Franceschini_2017}   & $0.96 \pm 0.05 \pm 0.06$                   & $0.8\sigma$ & $18.9\sigma$ & 356 \\
\citet{2021MNRAS.507.5144S} & $0.99 \pm 0.05 \pm 0.06$                   & $0.2\sigma$ & $19.1\sigma$ & 365 \\
\citet{Finke_2022}          & $1.04 \pm 0.05 \pm 0.07$                   & $0.9\sigma$ & $18.4\sigma$ & 339 \\
\hline
\end{tabular}
\tablecomments{The first uncertainty on $\alpha_\text{best}$ is statistical and the second is systematic. $R^{\sigma}(\alpha = \alpha')$ is the rejection significance of the hypothesis $\alpha = \alpha'$.  The Test Statistic compares the log-likelihood for the null hypothesis $\alpha = 0$ with that for the best-fit value $\alpha = \alpha_{\text{best}}$.}
\end{table*}

\begin{figure}[ht!]
\plotone{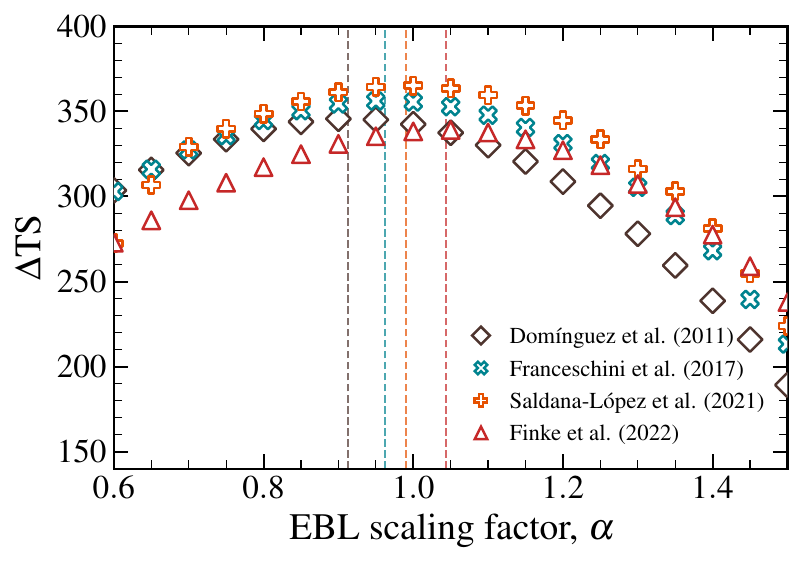}
\caption{Stacked $\Delta\mathrm{TS}$ as a function of the EBL scaling factor $\alpha$ for the four models used in the optical depth measurement: \citet{2011MNRAS.410.2556D}, \citet{Franceschini_2017}, \citet{2021MNRAS.507.5144S}, and \citet{Finke_2022}. Vertical dashed lines mark the best-fit $\alpha$ for each model.}
\label{fig:ts_alpha}
\end{figure}

The model that showed the best agreement with the data was that of \citet{2021MNRAS.507.5144S}, with $\alpha_{\text{best}}=0.99 \pm 0.05 \pm0.06$ closest to unity. This result supports the improved fidelity of the newer model over its predecessor, \citet{2011MNRAS.410.2556D}, which tended to favor lower normalization values in our dataset and appears to be less consistent with the current $\gamma$-ray observations. Within their respective uncertainties, the normalization factors inferred here are also consistent with the EBL intensities preferred by \citet{2026Banerjee..EBL} from GeV optical depths, indicating a coherent picture across energy bands. This implies that modern EBL models that already match galaxy counts and direct measurements at $z=0$ also reproduce the TeV attenuation pattern within $\lesssim 10\%$ in normalization up to $z \sim 0.94$. In particular, the close agreement with the \citet{2021MNRAS.507.5144S} model, which was calibrated on galaxy luminosity densities and SEDs, shows that the TeV data do not require any substantial additional diffuse component on top of the resolved galaxy light. Nevertheless, none of the EBL models tested show significant deviations from $\alpha=1$, indicating that the current Cherenkov telescope data set lacks the precision required to distinguish between them, with systematic uncertainties dominating the measurement. 

We note that some values in Table~\ref{tab:optical_depth} and Table~\ref{tab:ebl_density} exhibit highly asymmetric uncertainties. This behavior naturally arises from the highly non-linear transformation between the scale factor $\alpha$ and the derived optical depth within the profile likelihood, particularly near the edges of our sensitivity bounds. To illustrate the fit quality visually, Figure~\ref{fig:ts_alpha} shows the stacked $\Delta\mathrm{TS}$ profiles as a function of the EBL scaling factor $\alpha$ for four of the seven EBL models tested, selected for visual clarity. All four models exhibit broad, well-defined maxima near $\alpha \approx 1$, confirming that the fiducial EBL normalizations are consistent with the observed $\gamma$-ray attenuation. The vertical dashed lines indicate the best-fit $\alpha$ for each model, corresponding to the values reported in Table~\ref{tab:ebl_density}. The remaining three models show similar behavior.

\subsection{EBL Reconstructions}
\begin{figure*}[ht!]
\includegraphics[width=\textwidth]{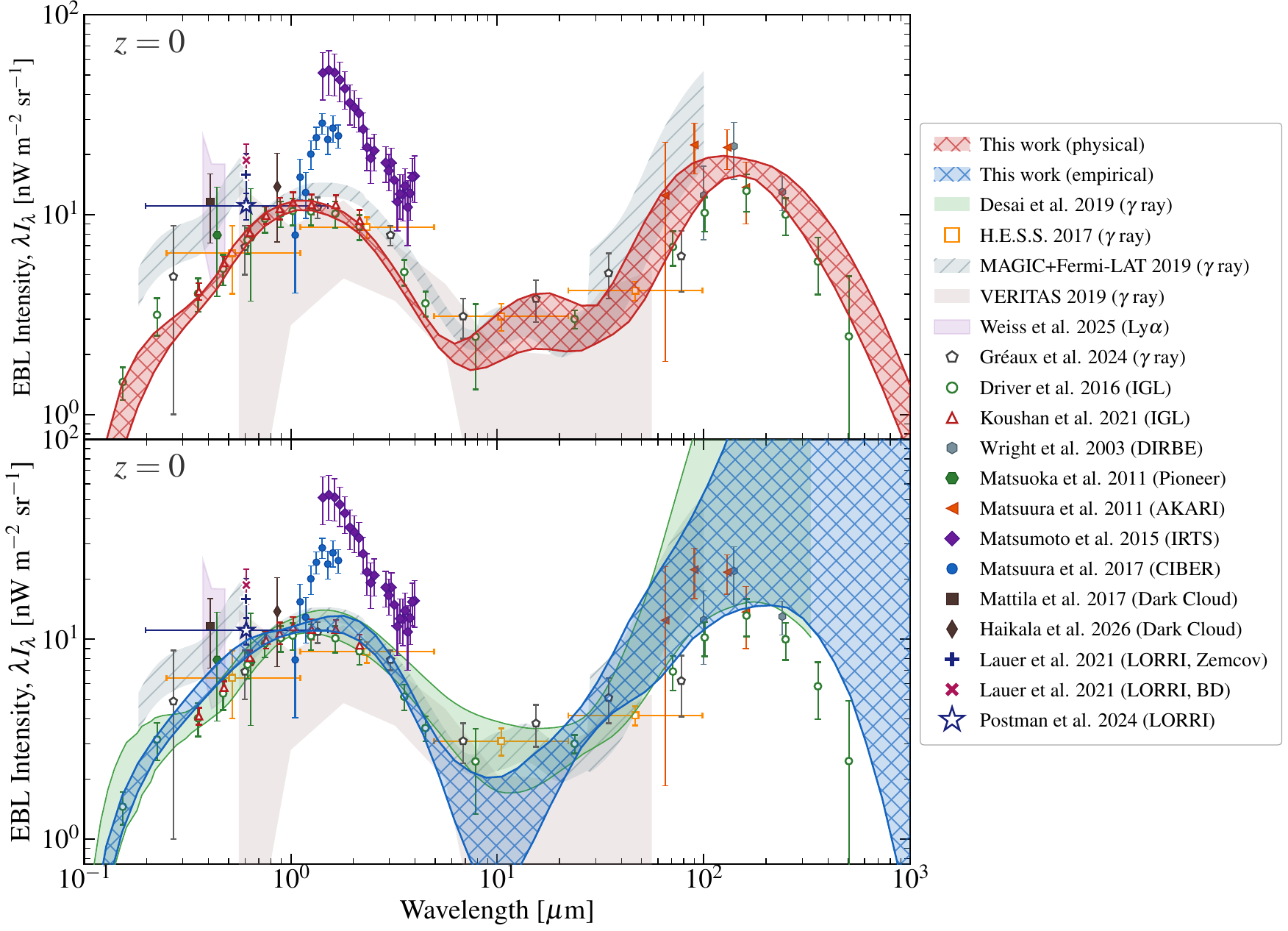}
\caption{Spectral energy distribution of the EBL at $z = 0$. \textit{Top panel}: reconstruction from our physically motivated EBL model (red shaded region denoting the 68\% confidence interval). \textit{Bottom panel}: empirical reconstruction (blue shaded region), shown together with the result of \citet[][green band]{2019ApJ...874L...7D}. In both panels, $\gamma$-ray-based constraints from H.E.S.S.~\citep{2017A&A...606A..59H}, MAGIC+\textit{Fermi}-LAT~\citep{2019MNRAS.486.4233A}, and VERITAS~\citep{Abeysekara_2019} are shown for comparison, together with the individual estimates from \citet{Greaux_2024}. Integrated galaxy light measurements from \citet{driver2016measurements} and \citet{Koushan_2021} are also included, as well as direct measurements and upper limits from Pioneer~\citep{Matsuoka_2011}, DIRBE~\citep{Wright_2001}, AKARI~\citep{Matsuura_2011}, IRTS~\citep{Matsumoto2015}, CIBER~\citep{Matsuura_2017}, LORRI~\citep{Lauer_2021, postman2024new}, Ly$\alpha$ forest constraints~\citep{Weiss_2025}, and the dark cloud method~\citep{Haikala_2026}, as indicated in the legend.}
\label{fig:ebl_intensity_combined}
\end{figure*}

Figure \ref{fig:ebl_intensity_combined} presents the results of both the physical framework and the empirical reconstruction, together with previous $\gamma$-ray-based studies, direct measurements, and the IGL results. The reconstructed EBL intensities and their residuals relative to IGL measurements are also explicitly reported in Table~\ref{tab:table3_bands}. The two independent reconstruction methods yield highly consistent EBL intensities, demonstrating that our results are robust to the choice of modeling approach. The derived spectrum is well-constrained across the optical to mid-infrared range, providing a precise VHE-anchored determination of the local EBL.

\begin{table*}[ht]
\caption{Mean IGL intensities and reconstructed EBL intensities in broad wavelength bands. $\lambda I_{\lambda}^{\rm IGL}$ denotes the mean integrated galaxy light, computed as simple averages within each wavelength band from the measurements compiled in \citet{driver2016measurements} and \citet{Koushan_2021}. The reconstructed EBL intensities from the empirical and physical models are denoted $\lambda I_{\lambda}^{\rm Emp}$ and $\lambda I_{\lambda}^{\rm Phys}$, respectively. Residuals are defined as $\Delta_{\rm Emp} = \lambda I_{\lambda}^{\rm Emp} - \lambda I_{\lambda}^{\rm IGL}$ and $\Delta_{\rm Phys} = \lambda I_{\lambda}^{\rm Phys} - \lambda I_{\lambda}^{\rm IGL}$, and the relative deviations are $\Delta_{\rm Emp}/\lambda I_{\lambda}^{\rm IGL}$ and $\Delta_{\rm Phys}/\lambda I_{\lambda}^{\rm IGL}$.}
\label{tab:table3_bands}
\centering
\begin{tabular}{lccccccc}
\hline\hline
Wavelength ($\mu$m) &
$\lambda I_{\lambda}^{\rm IGL}$ &
$\lambda I_{\lambda}^{\rm Emp}$ &
$\lambda I_{\lambda}^{\rm Phys}$ &
$\Delta_{\rm Emp}$ &
$\Delta_{\rm Phys}$ &
$\Delta_{\rm Emp}/\lambda I_{\lambda}^{\rm IGL}$ (\%) &
$\Delta_{\rm Phys}/\lambda I_{\lambda}^{\rm IGL}$ (\%) \\
\hline
0.3--0.6
& $4.6 \pm 0.3$
& $6.1 \pm 0.7$
& $4.4 \pm 0.4$
& $+1.5 \pm 0.8$
& $-0.2 \pm 0.5$
& $+32 \pm 18$
& $-5 \pm 11$
\\
0.6--2
& $10.4 \pm 0.5$
& $10.9 \pm 0.7$
& $10.0 \pm 0.6$
& $+0.4 \pm 0.9$
& $-0.5 \pm 0.8$
& $+4 \pm 9$
& $-4 \pm 7$
\\
2--10
& $5.1 \pm 0.5$
& $4.8 \pm 0.7$
& $4.0 \pm 0.3$
& $-0.3 \pm 0.9$
& $-1.1 \pm 0.6$
& $-6 \pm 17$
& $-21 \pm 9$
\\
\hline
\end{tabular}
\end{table*}

\section{Discussion and Conclusions} \label{sec:discussion}

To assess the consistency between our $\gamma$-ray based reconstruction of the EBL intensity and other measurements, we compared our results at $z=0$ with both integrated IGL estimates and direct EBL measurements. Residuals were computed at each wavelength where measurements are available, together with the corresponding Z\text{-score}s and two-sided $p$-values. This approach allows us to test quantitatively whether the $\gamma$-ray-reconstructed EBL leaves any statistically significant residual that could be interpreted as an additional diffuse component beyond resolved galaxies.

\begin{figure*}[ht!]
\plotone{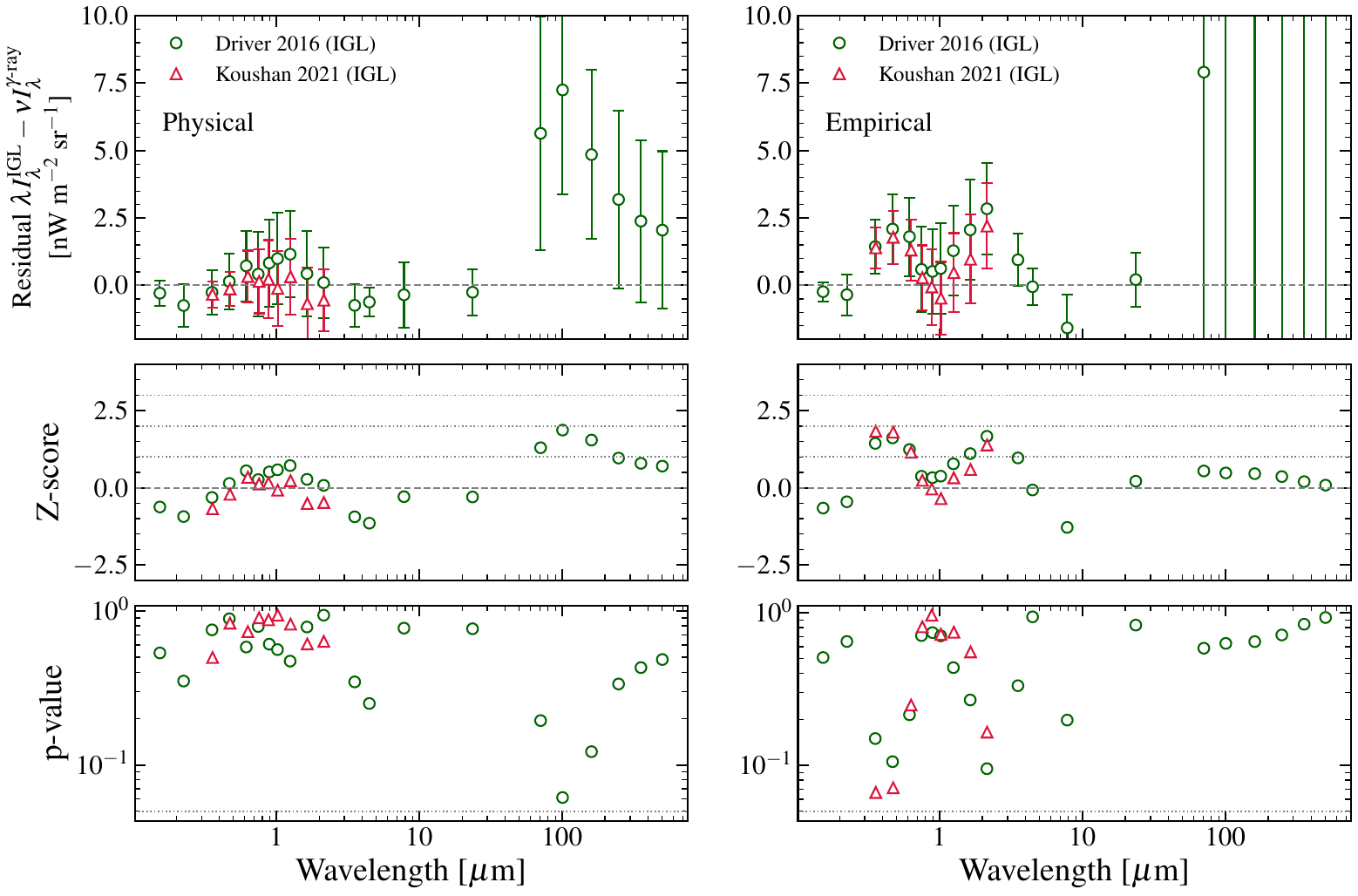}
\caption{Residual between our $\gamma$-ray derived extragalactic background light intensity and IGL measurements \citep{driver2016measurements, Koushan_2021} at $z=0$. Panels display, from top to bottom, the residual $R_i = \nu I_{\nu,i}^{\mathrm{IGL}} - \nu I_{\nu,i}^{\mathrm{\gamma ray}}$, the corresponding Z\text{-score}, and the two-sided $p$-value as a function of wavelength. Results are presented for both the empirical reconstruction and the physical one. The horizontal gray line in the bottom panel represents $p=0.05$.
}
\label{fig:residual_IGL}
\end{figure*}

Figure~\ref{fig:residual_IGL} shows the comparison with IGL measurements \citep{driver2016measurements, Koushan_2021}. Both the empirical and physical reconstructions are broadly consistent with the IGL across most wavelengths. Residuals generally remain within $\pm 2\sigma$, with only mild tensions in the optical regime ($\sim 0.3$--$0.5\,\mu$m), which are not statistically significant at the 95\% confidence level. In the $\sim 0.5$--$30$\,$\mu$m range, where $\gamma$-ray observations provide the strongest constraints, the agreement is particularly good. In this interval, residuals are smaller than \(2.8-3.0 \,\mathrm{nW\, m^{-2}\, sr^{-1}}\) and typically below \(33\%\) of the IGL intensity, indicating that any additional diffuse component must be at most a minor (\(\lesssim33\%\)) correction to the galaxy-count-based EBL.

Our findings are further contextualized by other recent approaches. \citet{Greaux_2024} recently measured the local ($z=0$) EBL over 0.18–120 $\mu$m by treating redshift evolution as a nuisance parameter. While we utilize a similar TeV dataset, our methodology differs by deriving optical depths $\tau(E_\gamma, z)$ to be combined with GeV measurements. This enables a joint reconstruction of both the EBL intensity and its redshift evolution.

\begin{figure*}[ht!]
\plotone{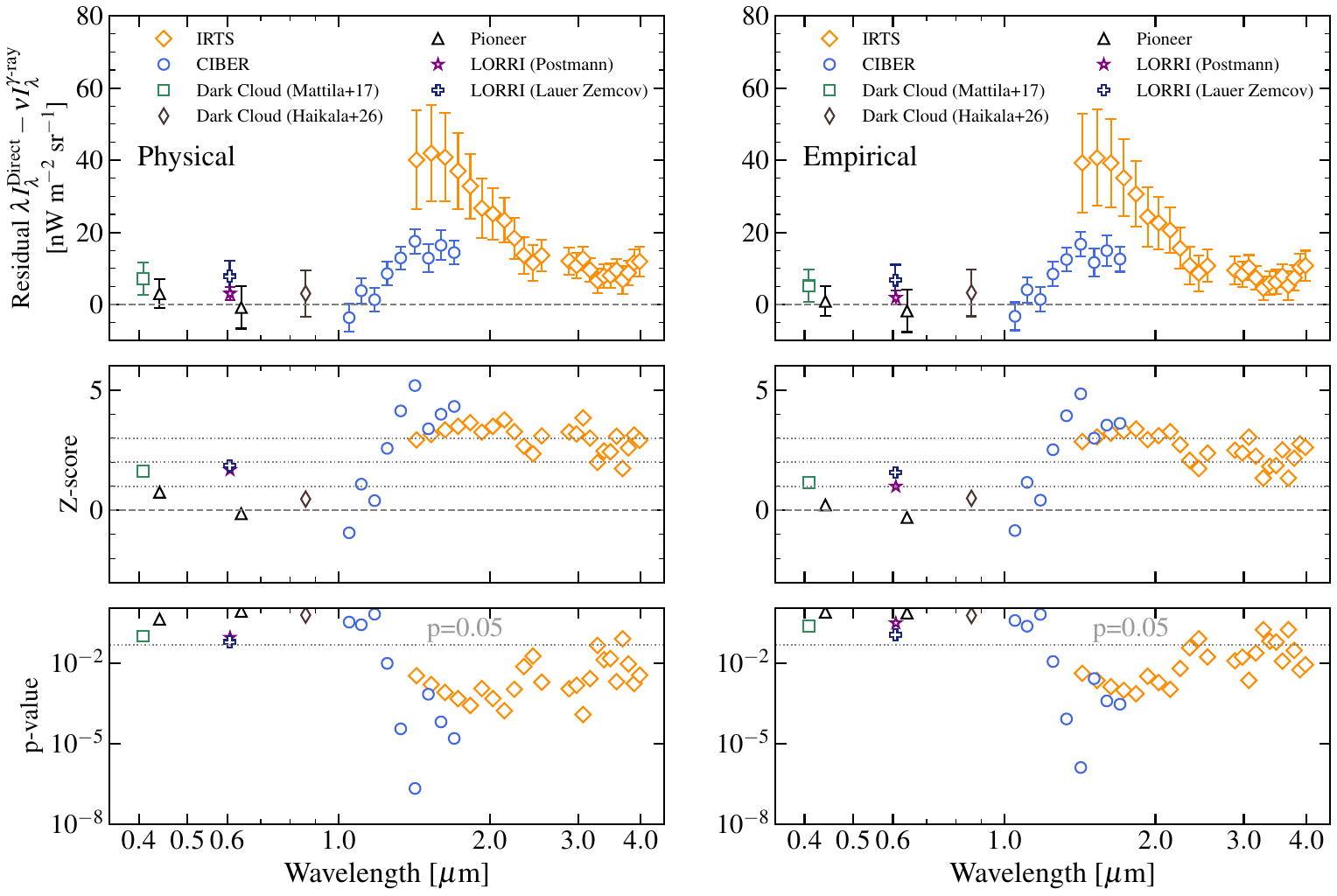}
\caption{Residual comparison between direct measurements of the EBL and $\gamma$-ray reconstructions. The left column shows results for the empirical reconstruction, and the right column shows the physical forward model. Panels display, from top to bottom, the residual $R_i = \nu I_{\nu,i}^{\mathrm{IGL}} - \nu I_{\nu,i}^{\mathrm{\gamma ray}}$, the corresponding Z\text{-score}, and the two-sided $p$-value as a function of wavelength. Markers indicate data from different instruments: IRTS (orange diamonds; \citealt{Matsumoto2015}), CIBER (blue circles; \citealt{Matsuura_2017}), Pioneer 10/11 (black triangles; \citealt{Matsuoka_2011}), New Horizons LORRI (purple stars; \citealt{postman2024new}, navy plus signs with Zemcov ZL subtraction; \citealt{Lauer_2021}), and the dark cloud method (green squares; \citealt{Mattila_2017}, brown diamonds; \citealt{Haikala_2026}) The horizontal gray line in the bottom panel represents $p=0.05$.}
\label{fig:residual_direct}
\end{figure*}

In Figure~\ref{fig:residual_direct}, we compare our results with direct EBL measurements. The most notable outcome is the agreement with the recent New Horizons LORRI results \citep{postman2024new}, obtained at large heliocentric distances where zodiacal light contamination is minimal. Both reconstructions yield residuals consistent with zero within $2\sigma$. Quantitatively, at optical wavelengths around \(0.4\text{ to }0.5\,\mu\mathrm{m}\), the difference between our reconstruction and the LORRI-based EBL measurement is smaller than \(27\%\) and statistically insignificant \((|z| \lesssim 2)\). Similar consistency is found with the dark cloud method \citep{Mattila_2017, Haikala_2026} and Pioneer 10/11 \citep{Matsuoka_2011}, further supporting the conclusion that the optical EBL is largely explained by resolved galaxy populations.

By contrast, substantial discrepancies are seen with the IRTS \citep{Matsumoto2015} and CIBER \citep{Matsuura_2017} measurements, where residuals often exceed $2\sigma$ and in some cases approach $4$--$5\sigma$. These datasets were obtained in the inner solar system, where zodiacal foregrounds are bright and variable, likely reflecting underestimated systematic uncertainties in the foreground modeling. While this is the most plausible explanation, exotic possibilities cannot be completely excluded; for example, \citet{2024PhRvD.110j3501P} explored an additional EBL contribution from the decay of dark matter consisting of axion-like particles (ALPs). However, interpreted literally as an additional macroscopic diffuse EBL component, the IRTS and CIBER excesses would require an increase of order $\sim 30$\% or more over the IGL in the near-IR. Our $\gamma$-ray based reconstruction, which is sensitive to the total EBL including any non-galaxy contribution, strongly disfavors such a large component. The $p$-values in Figure~\ref{fig:residual_direct} are well below $0.05$ in the corresponding bands, and the required EBL intensities are incompatible with our $\tau(E_\gamma, z)$ measurements at more than the $3\sigma$ level.

\begin{figure}[ht!]
\plotone{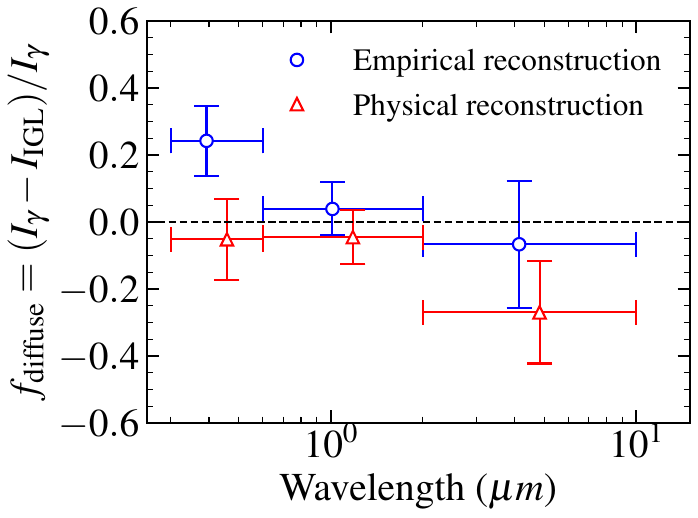}
\caption{Fraction of the EBL that may remain as a diffuse component in three broad wavelength bands, computed as $f_{\mathrm{diffuse}}=(I_{\gamma}-I_{\mathrm{IGL}})/I_{\gamma}$ from the reconstructed $\gamma$-ray EBL intensities (empirical and physical models) and the integrated galaxy light. Points are shown at the geometric mean wavelength of each band, with small horizontal offsets for clarity.}
\label{fig:f_diffuse}
\end{figure}

Combining the residuals across the \(0.8\text{ to }5\,\mu\mathrm{m}\) range, we obtain an approximate \(95\%\) upper limit on any additional diffuse component of \(\Delta I \lesssim 2.95 \,\mathrm{nW\, m^{-2}\, sr^{-1}}\), corresponding to \(\lesssim 24.1 \%\) of the IGL level in that band (see Fig.~\ref{fig:f_diffuse}). This is a bit more restrictive than the limits inferred in \citet{2026Banerjee..EBL} by comparing their EBL reconstruction with galaxy luminosity density data.

Taken together with the GeV-based analysis of \citet{2026Banerjee..EBL}, our results provide a coherent picture of the EBL. \citet{2026Banerjee..EBL} use updated \textit{Fermi}-LAT optical depths to trace the EBL intensity and the cosmic star formation history out to \(z \simeq 4\), while the present TeV-anchored study focuses on the local \((z = 0)\) EBL and on its agreement with galaxy counts and direct measurements. The consistency between the two reconstructions across overlapping wavelengths and redshifts supports a scenario in which known galaxy populations dominate the EBL over UV to mid IR wavelengths, with only limited room for additional diffuse light.

In summary, we combine an extensive set of TeV optical depths from STeVECat with GeV measurements from \citet{2018_science_fermi_ebl}. This physically motivated, TeV-anchored reconstruction is consistent with both integrated galaxy light and the most reliable direct measurements. By linking the observed attenuation to the cosmic star formation history, this analysis shows that resolved galaxies account for the bulk of the EBL in the optical and near-IR. Any additional diffuse component must therefore be modest.

\section{Acknowledgments}
J.B. gratefully acknowledges support from JSPS KAKENHI Grant No. 24KJ0545 and the JSPS Overseas Challenge Program for Young Researchers. J.B. also thanks UCM for hosting his six-month stay under the latter program, and A. Dom\'inguez for his warm welcome. A. Dom\'inguez is thankful for the support of Proyecto PID2021-126536OA-I00 funded by MCIN / AEI / 10.13039/501100011033. J.D.F. was supported by NASA through contract S-15633Y; the Office of Naval Research; and by a grant of computer time from the Department of Defense High Performance Computing Modernization Program at the Naval Research Laboratory. A.\ Desai was supported by an appointment to the NASA Postdoctoral Program at NASA Goddard Space Flight Center, administered by Oak Ridge Associated Universities under contract with NASA. A.B. and M.A. acknowledge NASA funding under contract 80NSSC22K1579.

\section{Data Availability}
The very-high-energy gamma-ray spectra analyzed in this study are publicly accessible through the STeVECat catalog \citep{greaux2024stevecat}. This database compiles data from ground-based Cherenkov telescopes, and the full repository can be accessed via Zenodo \citep{stevecat_zenodo}. The formatting of the STeVECat datasets aligns with the open-source standards established by Gamma-Cat and VTSCat. The VERITAS specific data products are maintained within the VTSCat repository \citep{vtscat_zenodo}. Additionally, the GeV optical depths utilized for the joint extragalactic background light reconstruction were derived from publicly available \textit{Fermi}-LAT data, as detailed by \citet{2018_science_fermi_ebl}.

\facilities{H.E.S.S., MAGIC, VERITAS, Fermi-LAT}

\software{Astropy \citep{astropy13, astropy18, astropy22}, 
          emcee \citep{emcee13}, 
          SciPy \citep{scipy20}}

%

\vspace{5mm}

\bibliography{merged.bib}{}
\bibliographystyle{aasjournal}



\end{document}